\documentclass{aa}  
\pdfoutput=1
\usepackage{graphicx}
\usepackage{amsmath}	
\usepackage{amssymb}	
\usepackage{relsize}
\usepackage{txfonts}
\usepackage[caption=false]{subfig}
\usepackage[colorlinks=true,linkcolor=blue,allcolors=blue]{hyperref}%
\usepackage[defaultcolor=red]{changes}
\usepackage{ulem}
\usepackage{soul}
\usepackage{gensymb}
\usepackage{appendix}

\begin{document} 

\title{Disk tilts in Milky Way–like galaxies and their link to mergers}

\subtitle{}

\author{Runsheng Cai$^{1,2}$\thanks{E-mail: cairunsheng@shao.ac.cn}, Ling Zhu$^1$\thanks{Corr author: lzhu@shao.ac.cn}, Jiang Chang$^3$, Lan Zhang$^4$, Xiangxiang Xue$^{4,5}$, Xi Kang$^6$ }

\institute{
Shanghai Astronomical Observatory, Chinese Academy of Sciences, 80 Nandan Road, Shanghai 200030, China
\and
School of Astronomy and Space Sciences, University of Chinese Academy of Sciences, No. 19A Yuquan Road, Beijing 100049, People’s Republic of China
\and
Purple Mountain Observatory, Chinese Academy of Sciences, Nanjing 210023, China
\and
National Astronomical Observatories, Chinese Academy of Sciences, Beijing 100101, China
\and
Institute for Frontiers in Astronomy and Astrophysics, Beijing Normal University, Beijing 102206, China
\and
Institute for Astronomy the School of Physics, Zhejiang University, 38 Zheda Road, Hangzhou, 310027, Zhejiang, China \\
 }

\date{Received; accepted}
   
\titlerunning{}
\authorrunning{Cai et al.}  
 
  \abstract
   {
   We investigated disk reorientation and its relationship to mergers in a sample of 93 Milky Way–mass galaxies drawn from the TNG50 simulations.  
From this sample, we identified 117 individual disk-tilt events with clearly defined start and end times, and we divided them into two classes based on whether the disk evolves smoothly or undergoes rapid disruption and regrowth during the tilt duration. For each tilt event, we linked the most massive merger that occurs in close temporal association with it. We focus on those tilt events occurring during phases of smooth disk evolution, which typically show tilt rates below $10^\circ$ per Gyr, can persist for 2-9 Gyr, and are generally connected to minor mergers. Roughly $40\%$ of these events reach tilt angles exceeding $30^\circ$, most often triggered by satellites whose masses are comparable to or less massive than the galaxy’s total stellar mass, merging on tangential orbits with orbital angular momentum that is retrograde or nearly perpendicular to the disk spin.  
The satellite consistently drives the disk tilt toward alignment with the satellite’s orbital spin; satellites that came in on retrograde orbits relative to the disk spin can thus induce large tilt angles and can produce an almost orthogonal configuration between the stellar disk and the satellite’s orbital plane as an intermediate phase. Owing to the comparatively slow tilt rates, the stellar disk can keep stable and continue to grow throughout the reorientation process. Consequently, galaxies that experience strong disk tilts can still maintain stellar disks that are as thin as those in systems without significant reorientation.
   }


   \maketitle

%-------------------------------------------------------------------

\section{Introduction}

 Galaxy growth via mergers is a fundamental aspect of hierarchical structure formation, as demonstrated by simulations that validate the crucial role of accretion events in building up galactic halos. The accretion of satellites can significantly change the angular momentum orientation of a halo, known as spin flip. Analysis of cosmological N-body simulations reveals that 35\% of dark matter haloes undergo a spin flip greater than $45\degree$, unrelated to major mergers, and spin orientation changes even more frequently in the inner halo \citep{2015spinflip}.  

The galaxy discs that are tightly bound by self gravity tilt along with the reorientation of the inner halo. The disc tilt is often attributed to torques from satellite mergers \citep{Binney1986, Ostriker1989,Bell2026}, as confirmed by numerical simulations \citep{Huang1997}. Beyond merger-driven effects, alternative mechanisms can produce a similar disc tilt, including the influence of non-spherical dark matter halo geometries \citep{Dubinski1992} and gas accretion processes \citep{Debattista2015, Debattista2013}.

In particular, direct N-body simulations show that a GSE-like merger \citep{Belokurov2018MNRAS.478..611B,Helmi2018Natur.563...85H} can effectively induce a disc tilt in Milky Way (MW)-like galaxies, which might eventually lead to a flip of the Galactic disc; with the tilt rate being sensitive to the geometry of merger orbits \citep{2022Dodge}.
Such disc tilt was also frequently observed in zoom-in MW-like hydrodynamic simulations \citep{Gomez2017a, Gomez2017b, Earp2017MNRAS.469.4095E, Dillamore2022, Dillamore2023}.
 GSE-like mergers are found in a few cases that indeed lead to significant disc tilt, and there is evidence that the ultimate orientation of the disc also depends on the orbital properties of the satellites \citep{Dillamore2022,Dillamore2023}. 
 Even low-mass satellites that penetrate the outer regions of a galaxy are found to significantly perturb and tilt a host galaxy disc \citep{Gomez2017b}, due not only to direct tidal perturbation but also to the generation of asymmetric features in the DM halo that can be efficiently transmitted to its inner regions. The disc tilt angle is thought to be correlated with the duration of close satellite interactions. The start of interactions is defined as the satellite's first crossing of the host's virial radius, and the end is defined as the satellite's final merger with the host \citep{Gomez2017a}. 

Disc tilt can significantly influence the evolution of Galactic structure. The theoretical work predicts that a tilted disc naturally develops a warp \citep{Binney1986}, as torques might cause the inner and outer rings to precess at different nutation rates \citep{Kazantzidis2009,Laporte2018,Wang2026ApJ..1006..160W}. 
However, it is not necessarily the origin of the warps in the MW disc. Gas accretion, the gravitational potential of a galaxy itself, and variations in the gravitational potential of the DM halo can generate warps \citep{Earp2017MNRAS.469.4095E,Earp2019,Sellwood2022,Deng2026,Johri2026}.

 Over longer timescales, disc tilting can reorient the disc by a substantial angle. A recent study indicates that the MW’s DM halo is orientated nearly perpendicular to the stellar disc, with its principal plane likely aligned with the satellite plane \citep{Zhu2026A&A...706A.193Z}. Simulations further suggest that the disc was initially co-aligned with the DM halo and satellite plane but has tilted over the past few Gyr at roughly $10^\circ-15^\circ$ per Gyr, ultimately yielding this vertical configuration \citep{Zhu2026A&A...706A.193Z}. Not long after, an ongoing disc tilting rate of $14\pm2^\circ\,\mathrm{Gyr}^{-1}$ around the Sun–Galactic Centre axis was directly measured from Gaia data \citep{Thomas2026}. A large disc tilt angle of $\gtrsim 90^\circ$ was also suggested to support the weak prograde rotational signal of the MW halo \citep{Batrakov2026}. 
  Such an ongoing tilt rate of $\sim 15^\circ$ per Gyr would also modify stellar orbits and stream morphologies, improving models of the Pal~5 stream \citep{2024NibauerStream}.

To gain further insight into how such a disc tilt could be induced, if it indeed occurred in the MW, and to further understand the specific configuration of the MW from a statistical standpoint. We investigate the frequency of disc tilts that occur in MW-like galaxies and especially how the disk tilt depends on the merger mass and orbital configuration of the associated satellite galaxies. We use MW–like galaxies from the TNG50 simulation \citep{2023arXiv230316217P} for this study.

This paper is structured as follows. In Section~\ref{SS:data}, we introduce the MW-like galaxies from TNG50, which we use in the following analysis, and define the basic parameters. In Section~\ref{SS:result}, we will introduce our results. We discuss in Section~\ref{SS:discussion} and conclude in Section~\ref{SS:conclusion}.

%. This simulation provides well-resolved galaxy structures while still yielding a relatively large sample of 198 MW- or M31-analogues. As a fully cosmological simulation, it incorporates key physical processes such as realistic merger histories, gas accretion, and galaxy–galaxy interactions. This allows for a more realistic study of disc tilts potentially associated with mergers, albeit in an environment where multiple mechanisms are simultaneously at play.

\begin{figure*}
\centering\includegraphics[width=6cm]{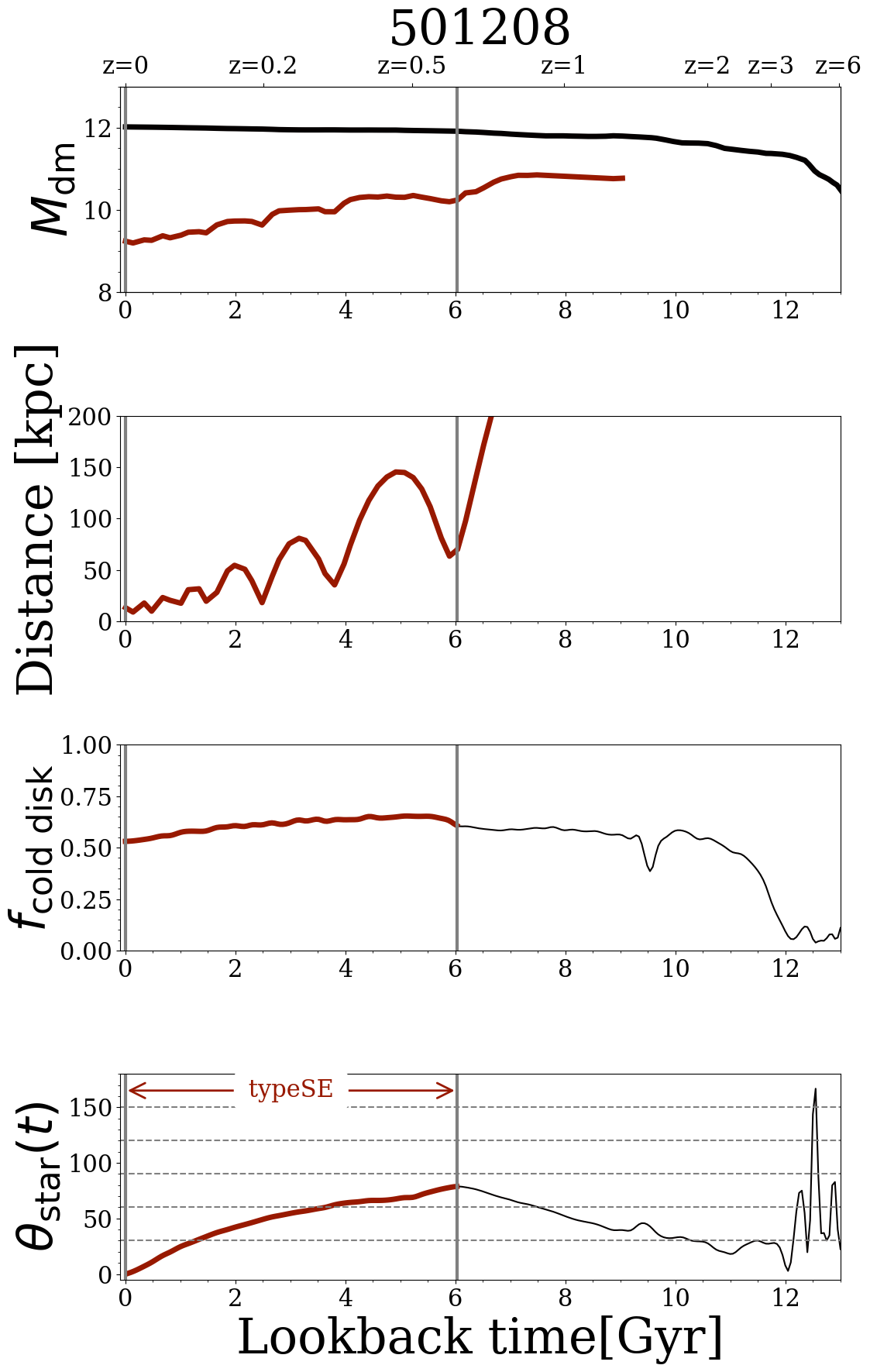}\centering\includegraphics[width=6cm]{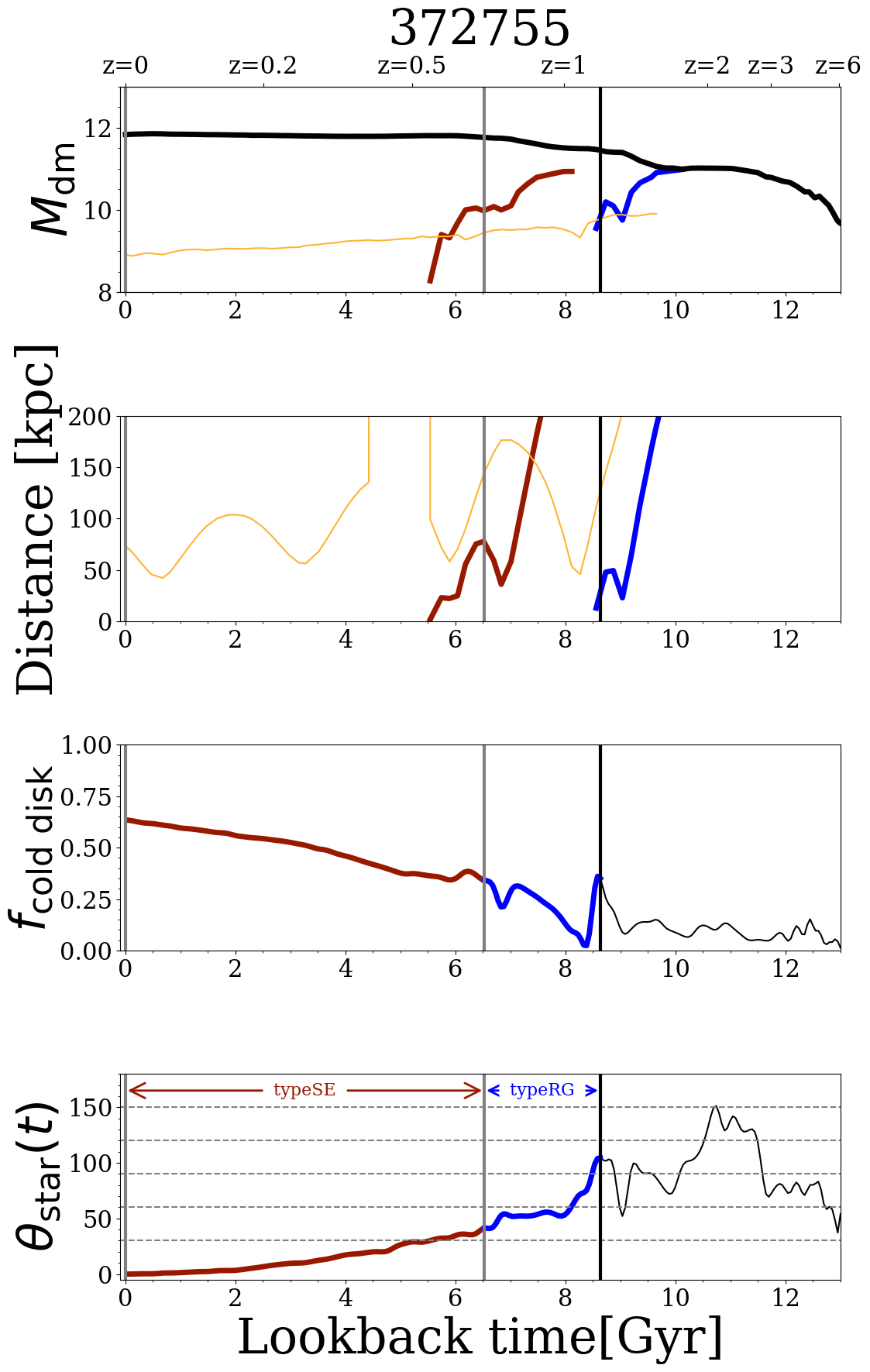}\centering\includegraphics[width=6cm]{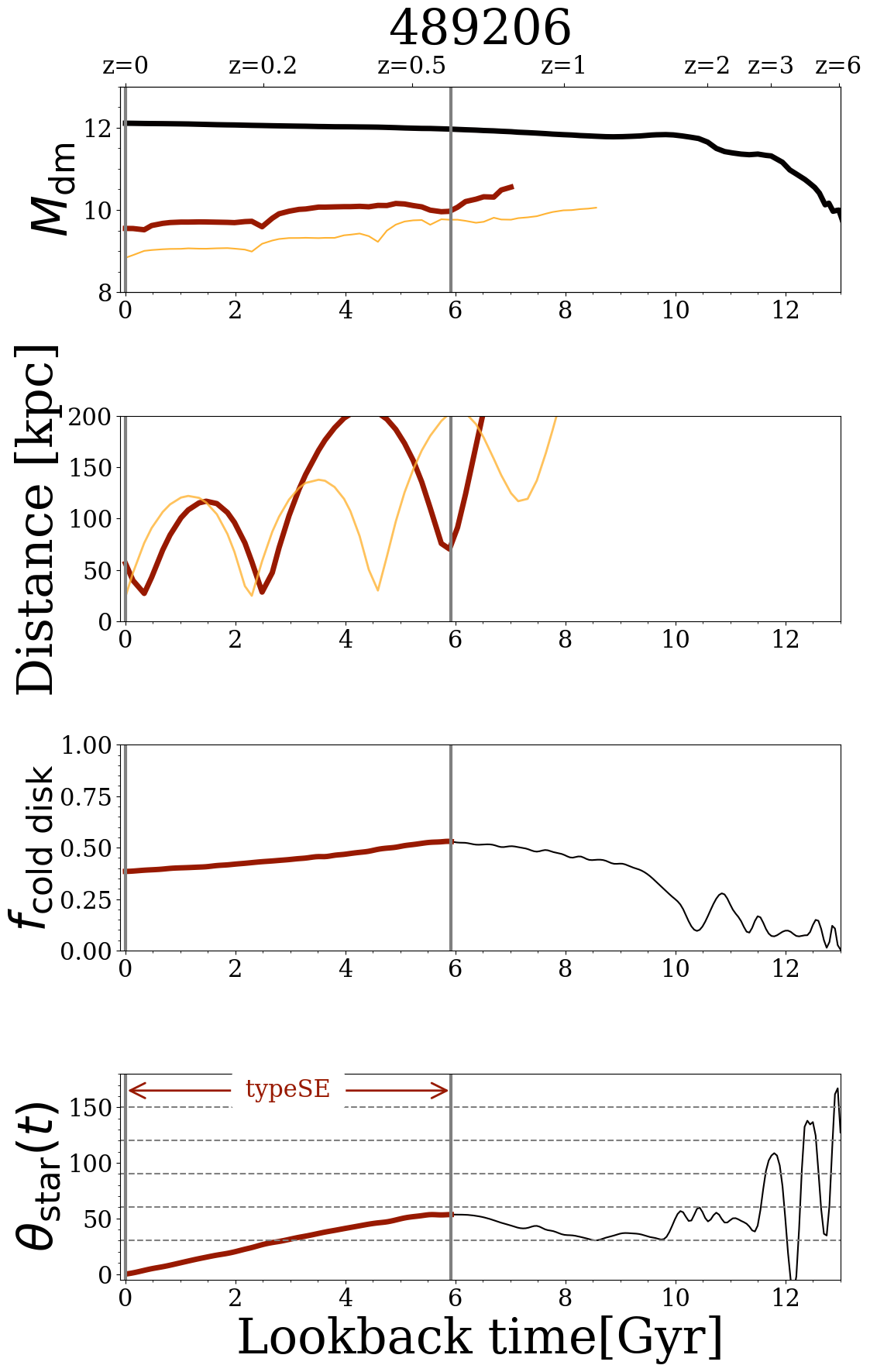}
\caption{Evolution history of TNG50 501208, 372755 and 489206. \textbf{Top}: Evolution of dark matter mass ($M_{\rm dm}$) over time. The black line represents the dark matter mass of the host galaxy, the thick red and blue lines indicate the dominant satellite we identified for the corresponding disk tilt event, the yellow lines are other massive satellites that likely also associated with the disk tilt events. \textbf{Second row}: orbits of the dominant satellites described as the evolution of its distance to the main galaxy. \textbf{Third row}: Evolution of the disk fraction ($f_{\rm cold\ disk}$). The red curves are identified as episodes with smooth disk evolution, while the blue is the episode with disk disruption or rapid growth. \textbf{Bottom}: Evolution of the disk spin orientation ($\theta_{\rm star}(t)$). The red/blue lines indicate the duration of the corresponding disk tilt events we finally identified, the vertical lines mark the starting point of the corresponding tilt events. }
\label{fig:time_region}
\end{figure*}

\begin{figure}
\centering\includegraphics[width=8cm]{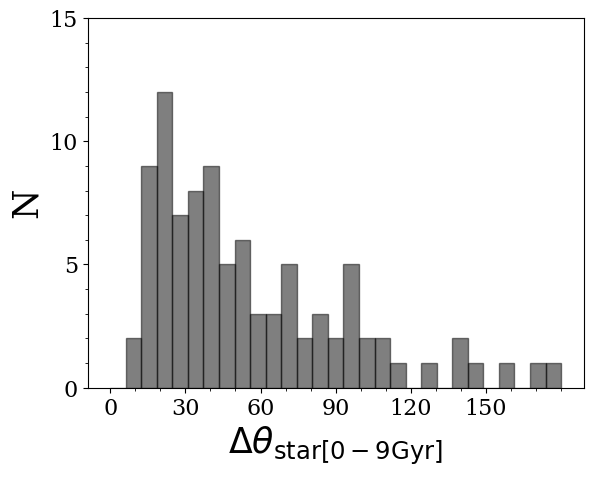}
\caption{The distribution of maximum angles the disk has tilted in the past 9 Gyr, $\Delta\theta_{\rm star}[0-9\,\rm Gyr]$, for all the 93 galaxies. }
\label{fig:Delta_theta_9Gyr}
\end{figure}

\begin{figure*}
\centering\includegraphics[width=18cm]{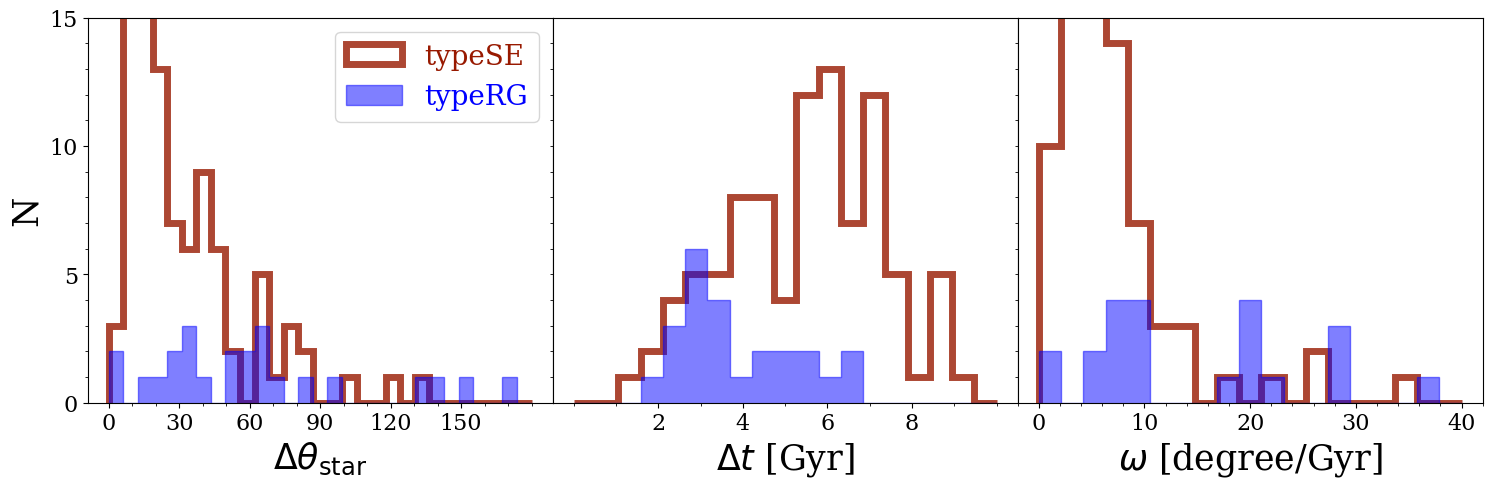}
\caption{The distribution of total tilt angle $\Delta\theta_{\rm star}$, tilt duration $\Delta t$, and tilt rate $\omega$ between the onset time $t_s$ and end time $t_e$, for the disk tilt events that occurred during smooth disk evolution (type SE) and during rapid growth disk (type RG) that identified from the 93 galaxies.}
\label{fig:theta_t_w_hist}
\end{figure*}

%---------------------------------------------------------------------
\section{Data}
\label{SS:data}
%---------------------------------------------------------------------

We use MW-like galaxies from TNG50 simulations \citep{Nelson2019ComAC...6....2N, Nelson2019MNRAS.490.3234N, Pillepich2019MNRAS.490.3196P} for our analysis. In this section, we first introduce the definition of some general parameters in simulations, then define the galaxy sample and disc tilt event in the sample.

\subsection{Merger tree}
\label{SS:Merger tree}
In the cosmological simulations, the merger tree is a fundamental tool that connects temporal snapshots of a simulation to reconstruct the evolutionary pathways of galaxies and their host dark matter halos. By tracking the unique particle IDs that constitute each halo \citep{Gill2004,Knollmann2009Halofinder}, it establishes “progenitor-descendant” links across time. Typically, a “majority particle inheritance” criterion (e.g., >50\%) defines continuity, while events where a descendant halo receives significant mass from multiple progenitors are marked as mergers \citep{Springel2001MNRAS,Dolag2009}. The resulting tree structure allows us to precisely identify the continuous growth history of any galaxy along its main branch, as well as the accretion of satellite galaxies via secondary branches. This serves not only as the chronology for studying mass assembly history but also as the definitive method for distinguishing central from satellite galaxies and for characterising merger events (e.g., major, minor, dry, wet) by their type and timing. 

Within the merger tree framework, satellite mass is the total mass of the satellite halo before it crosses the virial radius of its host halo and ceases to be a central galaxy \citep{Rodriguez-Gomez2019}. The mass ratio between the satellite and the central halo is the primary parameter for classifying merger types (e.g., >1:3 for a major merger) and predicting the degree of structural disruption to the central galaxy. In the top panels of Fig.~\ref{fig:time_region}, we illustrate the mass assembly history and the most massive mergers of three galaxies. 

\subsection{Stellar disk fraction}
\label{SS:Stellar disk fraction}

We use dynamical decomposition to define the structure of a galaxy, especially for the disc component. 
 We take the stellar particles of each galaxy at z=0 and align the X, Y, and Z axes as the three principal axes defined from the stellar mass distribution.

For each stellar particle, we compute its specific orbital angular momentum component ($j_{\rm Z}$) along the direction of the galaxy’s total angular momentum. 
This is compared to the angular momentum of a theoretical circular orbit at the same orbital radius ($j_{\rm c}(r)$), yielding $\lambda_{Z}$ = $j_{\rm Z}$/$j_{\rm c}(r)$. A stellar particle on a co-planar, circular orbit in an idealised thin disc has $\lambda_{Z}$ = 1, while a star in an isotropic, dispersion-supported spheroidal component (e.g., a bulge or stellar halo) has a symmetric $\lambda_{Z}$ distribution around zero, with a mean near zero. 

We define a dynamically cold disc with stars of circularity $\lambda_{Z}$ > 0.8; the disc fraction $f_{\rm cold\,disk}$ is defined as the mass fraction of these stars compared to the total stellar mass. In the third row of Fig. \ref{fig:time_region}, we show the $f_{\rm cold \,disk}$ changes over time. TNG50 501208 shows a cold disc that is increasing and maintaining a steady state, while TNG50 372755 shows $f_{\rm cold \, disk}$ near zero at about 8.5 Gyr due to a major merger and rebuilding after the end of the major merger process. The variation of $f_{\rm cold \, disk}$ is sensitive to disc disruption and disc rebuilding during and after the merger.

%------------------------------------
\subsection{Galaxy sample}
\label{SS:Galaxy sample}
%------------------------------------
We take 198 MW-like or M31-like galaxies \citep{Pillepich2024MNRAS.535.1721P} from the TNG50 simulations \citep{Nelson2019MNRAS.490.3234N,
Nelson2019ComAC...6....2N,  Pillepich2019MNRAS.490.3196P} and perform further selections to enhance their comparability with the MW, following \citet{Zhu2026A&A...706A.193Z}. 
We calculated the flattening of the stellar component $q_{\rm star} = Z/X$ of each simulation and defined the disc fraction $f_{\rm disc}$ by considering stars with circularity $\lambda_z>0.5$ as part of the disc. During the sample selection, we included some stars on dynamically warm orbits to make the disc comparable to the Milky Way disc, which was defined by fitting exponential profiles to the stellar number density distribution \citep{Grand2017MNRAS.467..179G}, rather than using the dynamically cold disc. The Milky Way has $f_{\rm disc} \sim 0.7$ and global $q_{\rm star} \sim 0.15$ \citep{Bland-Hawthorn2016}. We selected galaxies using a slightly loose threshold with $f_{\rm disc} > 0.6$ and $q_{\rm star} < 0.3$ to retain a relatively large sample. This selection removed galaxies that have experienced relatively recent major mergers, which create a more prominent bulge or stellar halo than that of the Milky Way. We further removed a few galaxies with ongoing major mergers that have not stabilised their stellar components. This results in 93 galaxies in our sample.

%-----------------------------
\subsection{Identification of disk tilt event}
\label{SS:classify}
%-------------------------------

We define the spin direction of the galaxy by taking all the stars within 20 kpc from the galactic centre, denoted as $L_{\rm main}$, to distinguish them from the satellite galaxies that will be mentioned later. All the galaxies in our sample are disc galaxies like the MW. The spin direction of $L_{\rm main}$ defined in this way is a good representation, and we take it as the disc spin throughout the paper.

The orientation of disc spin varies over time. We quantify the disc orientation during its lifetime by $\theta_{\rm star}(t)$ with fixed $\theta_{\rm star}(t=0) = 0$, where $t$ is the lookback time. The orientation of $L_{\rm main}$ at t=0 for each galaxy is set as the zero reference. The evolution of $\theta_{\rm star}(t)$ for three typical cases is shown in the bottom panels of Fig. \ref{fig:time_region}. 

We aim to identify disc-tilt events in which the disc keeps tilting in the same direction for a comparatively long interval. For each galaxy, we inspect the $\theta_{\rm star}(t)$ curve and locate all turning points where $d\theta_{\rm disk}(t)/dt=0$ occurs within $0<t<9$ Gyr. As illustrated in Fig.~\ref{fig:time_region}, galaxies such as TNG50 501208 and TNG50 489206 show an unambiguous turning point. In other cases (e.g., 372755), where $\theta_{\rm star}(t)$ shows some fluctuation at $t \gtrsim 8$ Gyr, we choose the turning point that is followed by a relatively smooth trend.

We take this turning point to define the start time of the tilt event, $t_s$. In most systems, the tilting persists to the present, so we set the end time to $t_e=0$. For a small number of galaxies, the tilting slows markedly and $\theta_{\rm star}(t)$ flattens before $t=0$; in those cases, we identify a subsequent turning point and adopt it as $t_e$.

Next, we examine how the cold disc fraction $f_{\rm cold\, disk}$ changes between $t_s$ and $t_e$. To characterize how smooth the evolution of $f_{\rm cold\, disk}$ is for each galaxy, we compute the standard deviation $\sigma$ of $f_{\rm cold\, disk}$ in a 1.5 Gyr rolling window, along with the disc growth rate per Gyr. We consider the disc to evolve smoothly over periods where $\sigma(f_{\rm cold\, disk})<0.02$ and the growth rate is $<0.3$ per Gyr, indicating that the disc is neither strongly disrupted nor undergoing rapid growth.

Out of 93 MW-like galaxies, 69 satisfy these smooth-evolution criteria throughout the full interval from $t_s$ to $t_e$ (e.g., TNG50 501208 and 489206 in Fig.1). We classify the disc tilt over this duration as an event with smooth disc evolution (Type SE). The remaining 24 galaxies (e.g., TNG50 372755 in the middle panel of Fig.~\ref{fig:time_region}) exhibit a rapid change in cold disc fraction early on, followed by a long phase of smooth evolution. For these systems, we split the history into multiple events: a Rapid Growth (Type RG) event spanning the disruption/rapid growth phase of the disc, followed by a Type SE event covering only the subsequent disc tilt phase with a smooth evolution of the disc. The boundary between them is set at the time when the $f_{\rm cold\, disk}$ growth curve transitions from fluctuating to smooth, using the same smoothness criteria as above. This transition time is defined as the ending time $t_e$ for the Type RG event and as the starting time $t_s$ for the following Type SE event.

From the 93 galaxies, we identified 93 type SE and 24 type RG disc tilt events. All type SE events occurred during a period of smooth disc evolution, without disc disruption or rapid growth, while type RG events occurred with significant disc fluctuations. For each disc tilt event, we have clearly identified its starting time $t_s$ and ending time $t_e$. 

We then cheque all the satellite galaxies that have been orbiting around at the starting time of the disc tilt $t_s$. We take the one with the maximum mass at $t_s$, denoted as $M_{\rm max-sat, t_s}$, as the dominant satellite likely associated with the disc tilt event. 
For 45 of the 93 type SE events, there are no other satellites with a mass larger than $0.5 M_{\rm max-sat, t_s}$ at $t_s$; we thus expect the dominant satellite selected to have major effects on the disc tilt event, as in the case of TNG50 501208. For the remaining 48 events, there is at least one other satellite that is more massive than $0.5 M_{\rm max-sat, t_s}$ orbiting around at $t_s$, as in the case of TNG50 489206; other satellites might also have significant effects on the disc tilt.

\begin{figure}
\centering\includegraphics[width=9cm]{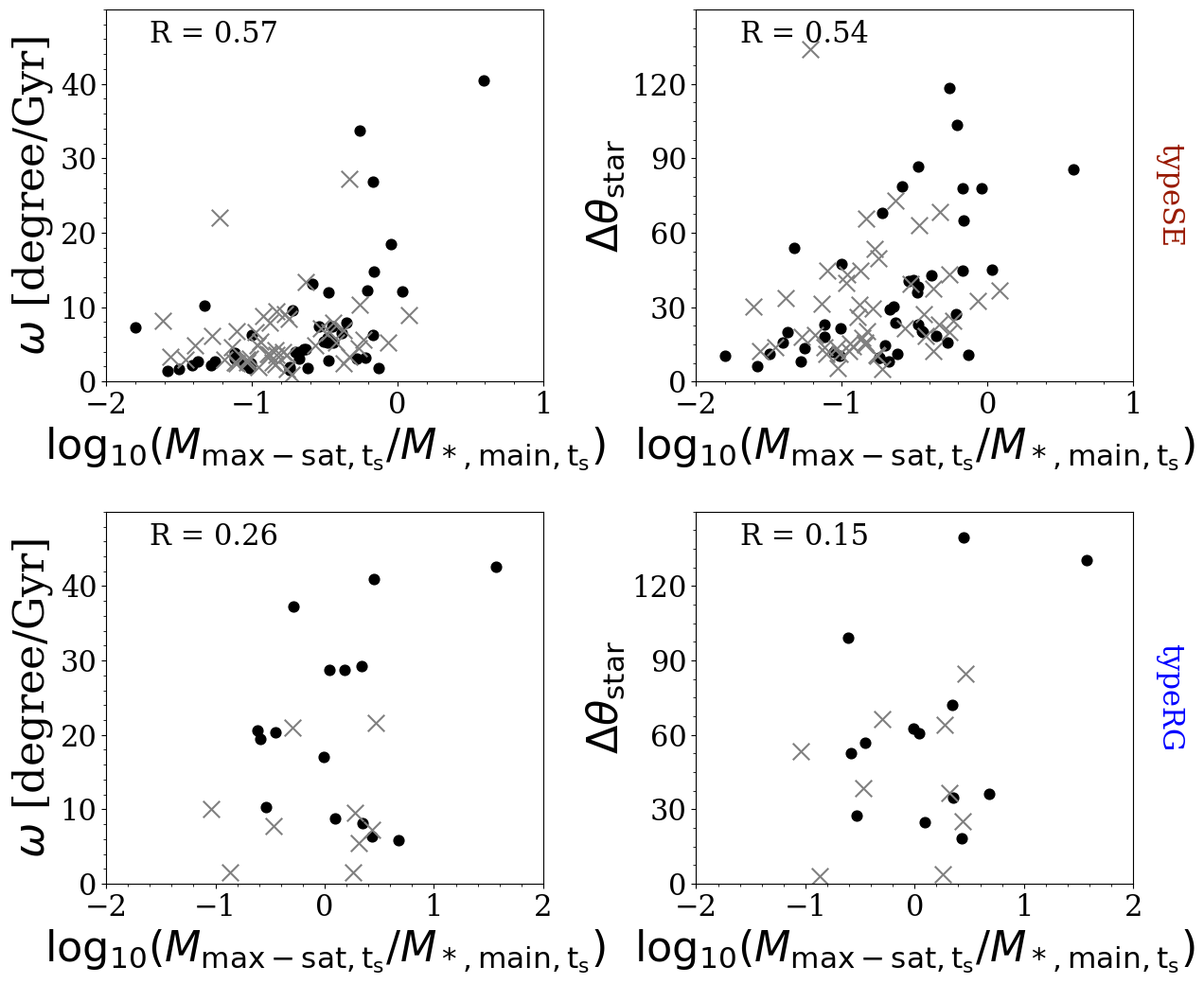}
\caption{Correlation between the mass ratio of the most massive associated merger to the disc ($M_{\rm max-sat, t_s}/M_{\rm *, main, t_s}$) versus tilt angle ($\Delta \theta_{\rm star}$, left) and tilt rate ($\omega$, right). Black dots denote hosts with one clearly dominant satellite during the tilt episode. Grey crosses indicate systems with multiple associated satellites with comparable mass; for which we add all satellite mass which is larger than 0.5 times of the most massive satellite. The top and bottom rows correspond to type SE and type RG events, respectively. The Pearson correlation coefficient $\mathcal{R}$ is denoted in each panel.
}
\label{fig:mass_ratio_in_ts}
\end{figure}

\begin{figure}
\centering\includegraphics[width=8cm]{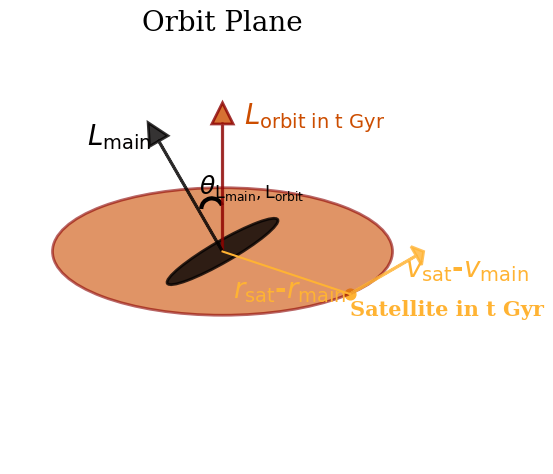}
\caption{Illustration of the merger orbital configuration at the starting point of disk tilt event $t_s$. The black and red disks represent the host galaxy plane and the orbital plane, respectively. The black and red arrows indicate the corresponding spin direction, we define $\theta_{\rm L_{main},L_{orbit}}$ as the angle between them. The yellow point marks the position of the satellite galaxy, while the yellow line and arrow represent the position vector $r_{\rm sat} - r_{\rm main}$ and velocity vector $v_{\rm sat} - v_{\rm main}$ of the satellite, respectively, we define $\theta_{r,v}$ as the angle between them.
}
\label{fig:show_nuta_prece_angle}
\end{figure}

\begin{figure}
\centering\includegraphics[width=8cm]{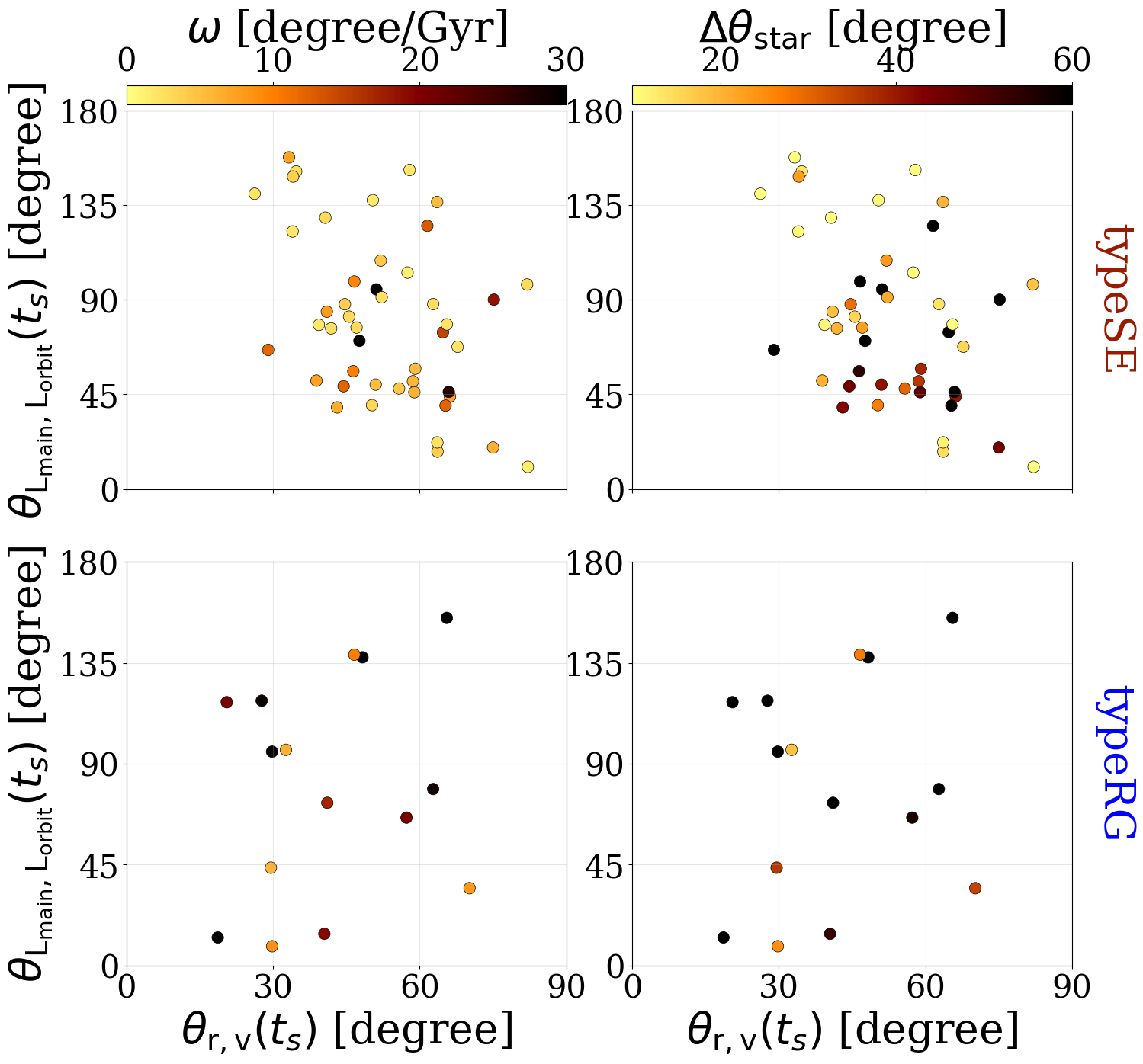}
\caption{The dependence of the disk tilt rate $\omega$ and tilt angle $\Delta \theta_{\rm star}$ on the merger orbital configuration of the most massive satellite $\theta_{r,v}(t_s)$ and $\theta_{\rm L_{main},L_{orbit}}(t_s)$. Most type SE events that show relatively rapid tilting ($\omega > 10^\circ$/Gyr or $\Delta\theta_{\rm star}>50^\circ$) are induced by satellites on tangential orbits with $\theta_{\rm r,v} \sim 60^\circ$, and with orbital angular momentum that is approximately vertical relative to the disk spin, with $45^\circ\lesssim  \theta_{\rm L_{main},L_{orbit}}\lesssim 135^\circ$.}
\label{fig:orbit_para_color_thetaw}
\end{figure}

\begin{figure}
\centering\includegraphics[width=7cm]{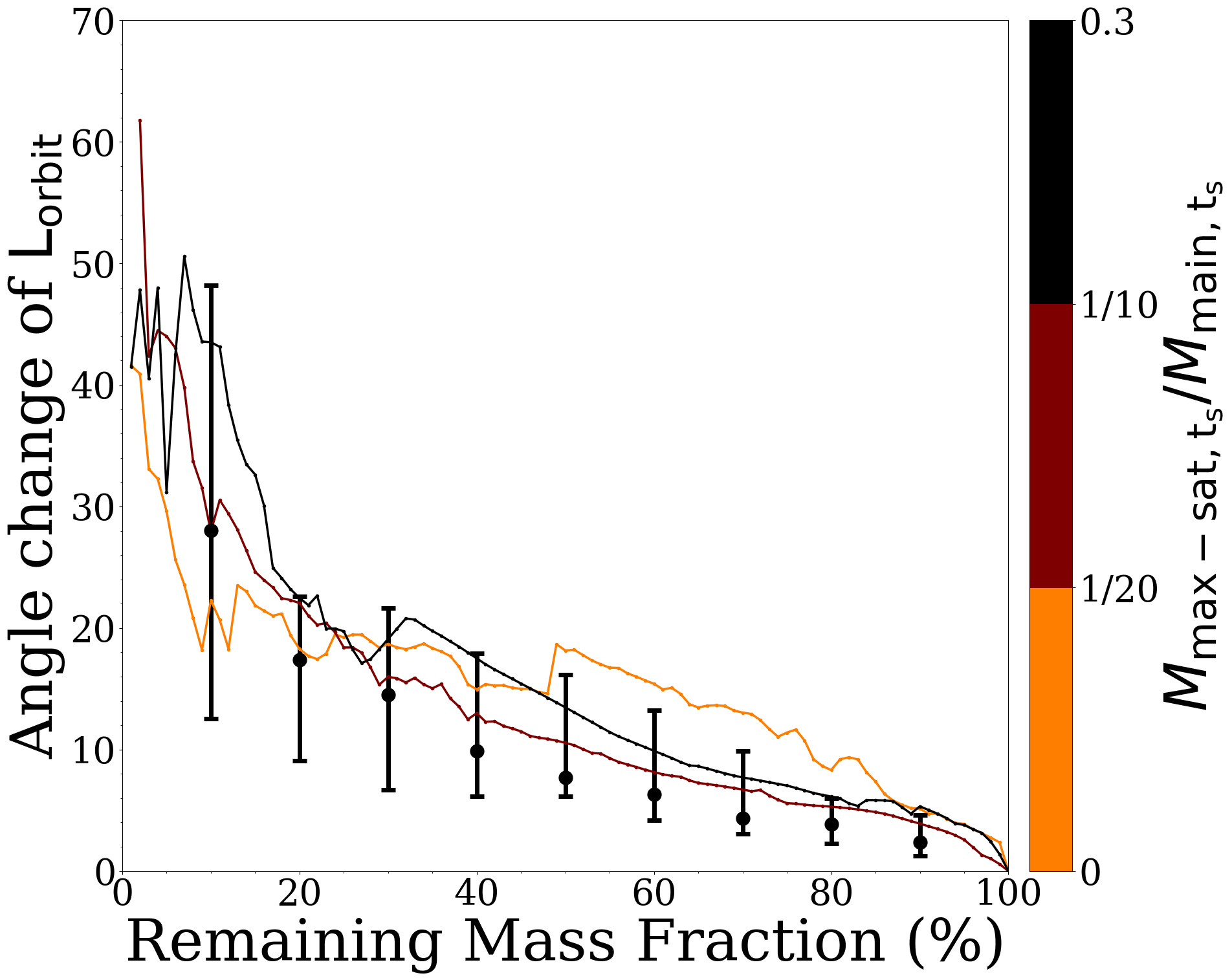}
\caption{The remaining mass fraction of satellite versus the change of its orbital spin direction ($L_{\rm orbit}$). The dots with error bars are the mean and scatter of all the satellites identified as dominant for the type SE events. The colored lines are the mean of satellites with different merger mass ratios as indicated by the colorbar. The merging satellites keep the direction of their orbital angular momentum without significant change ($\lesssim 20^\circ$) until they have more than $\sim 80\%$ of their mass disrupted (remaining mass fraction < $20\%$).}
\label{fig:RemainMass_Lorbitchange}
\end{figure}

\begin{figure*}
\centering\includegraphics[width=14cm]{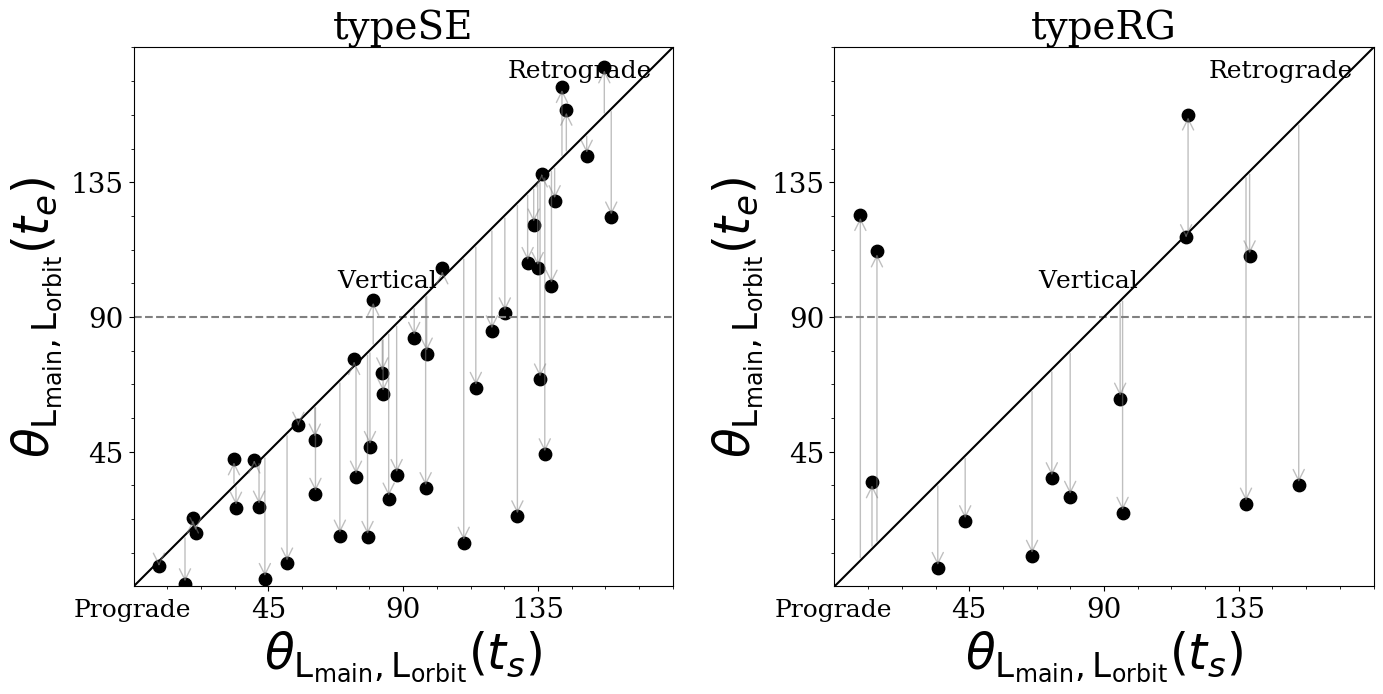}
\caption{The evolution of $\theta_{\rm L_{main},L_{orbit}}$ from the start of the tilt episode to its end. We only consider events with obviously one dominant satellite identified, showing type SE in the left panel and type RG in the right panel. Each point corresponds to one tilt event. The arrow marks the direction of evolution throughout the tilt episode. For all type SE events that display a large variation in $\theta_{\rm L_{main},L_{orbit}}$, the tilt concludes with a smaller value of $\theta_{\rm L_{main},L_{orbit}}(t_e)$; that is, the disc angular momentum systematically evolves toward alignment with the satellite’s orbital angular momentum at $t_{\rm s}$. }
\label{fig:theta_Lmainorbit_ts_te}
\end{figure*}

%---------------------------------------------------
\section{Result}
\label{SS:result}
%---------------------------------------------------
\subsection{A global view of disk tilt in MW-like galaxies}
\label{SS:statistics_all_galaxy}

We first present an overview of the disc tilt in the sample of 93 MW-like galaxies. We calculate the maximum change in disc orientation that has occurred over the past 9 Gyr relative to the disc at z=0. As shown in Fig. \ref{fig:Delta_theta_9Gyr}, 64 of our 93 galaxies have disc orientations that have changed by $>30\degree$, and 41 of these by $>50\degree$. 

With the two types of disc tilt identified and with clean start and end times introduced in Section \ref{SS:classify}, we quantify the disc tilt types with their tilt angle $\Delta\theta_{\rm star}$ = $\theta_{\rm star}(t_s) - \theta_{\rm star}(t_e)$, the time duration of the event $\Delta t$ = $t_s-t_e$, and the average tilt rate during the event $\omega =\Delta\theta_{\rm star} / \Delta t$. 

As shown in Fig~\ref{fig:theta_t_w_hist}, the type SE events usually have a lower tilt rate, mostly below $10^\circ$ per Gyr, yet there are about ten above. However, they have a relatively long duration of between 2 and 9 Gyr. As a result, 39 of the 93 type SE events have tilted at an angle $>30\degree$, and 16 at $>50\degree$. In contrast, type RG events generally have shorter durations and a higher tilt rate, with 18 of the 24 events having tilted at an angle $>30\degree$, and 14 at $>50\degree$ during the event.

%---------------
\subsection{The effects of merger mass on disk tilt}
\label{SS:mergerratio_effect_on_tilt}
%--------------

We have identified the dominant satellite associated with each disc tilt event, denoted as $M_{\rm max-sat, t_s}$. In systems such as TNG50 501208, at time $t_s$, there is obviously one dominant satellite; any additional satellites present in orbits around the galaxy are far less massive. In contrast, some systems experience several mergers of comparable mass in close temporal succession, as shown in TNG50 489206. In those cases, the disc tilt may result from the combined torques exerted by these multiple mergers. We flag such events when there is at least one additional satellite whose mass at $t_s$ exceeds $0.5 M_{\rm max-sat, t_s}$. 

In Fig.~\ref{fig:mass_ratio_in_ts}, we present how the mass ratio $M_{\rm max-sat, t_s}/ M_{\rm *, main, t_s}$ correlates with the tilt angle $\Delta\theta_{\rm star}$ and the tilt rate $\omega$. Here, the mass ratio is defined as the total mass of the dominant satellite ($M_{\rm max-sat, t_s}$) divided by the stellar mass of the main galaxy at the beginning of the tilt episode, $M_{\rm *, main, t_s}$.
For those with multiple massive satellites associated with $t_s$, we use the summation of all satellite masses greater than $>0.5 M_{\rm max-sat, t_s}$.

In general, we find that a larger $M_{\rm max-sat, t_s}$/$M_{\rm *, main, t_s}$ yields a greater tilt angle and higher tilt rate across type SE.
 The Pearson correlation coefficients between $M_{\rm max-sat, t_s}$/$M_{\rm *, main, t_s}$ and $\omega$ ($\Delta \theta_{\rm star}$) are 0.57 (0.54) for type SE. Most type SE events involve satellites whose total mass is far below that of the host galaxy’s stellar mass. By contrast, events with comparatively high tilt rates ($\omega > 10^\circ$/Gyr) tend to be associated with satellites whose mass is similar to the stellar mass of the main galaxy at that time, i.e., $M_{\rm max-sat, t_s}$/$M_{\rm *, main, t_s} \sim 1$. Note that here we are comparing the total mass of the satellite to the stellar mass of the main galaxy. These satellites are mostly still minor mergers compared to the total mass of the main galaxy (see Appendix Fig. \ref{fig:all_mass_ratio_in_ts}).

In contrast, type RG events—where the disc is disrupted—are typically triggered by more massive mergers, with $M_{\rm max-sat, t_s}$/$M_{\rm *, main, t_s} \sim 1-10$. A few of them are major mergers, but most are still minor mergers when considering their total mass ratio (see Appendix Fig.~\ref{fig:all_mass_ratio_in_ts}). Some of these encounters completely destroy the disc. In such cases, the change in disc orientation is tied to the disc’s disruption and subsequent reconstruction, rather than being directly linked to torques from the satellite’s mass. Consequently, there is no correlation between $M_{\rm max-sat, t_s}$/$M_{\rm *, main, t_s}$ and either the tilt rate or the tilt angle.

%---------------------------------------------------
\subsection{The effects of merger orbit configuration on disk tilt}
\label{SS:mergerorbit_effect_on_tilt}

After identifying the dominant merger that might have induced the disc tilt event, we further cheque the merger orbit and its effect on the tilt rate (angle). Their dependence on the merger angles is particularly evident in cases where one dominant satellite could be identified. In the following analysis, we only use type SE and type RG events with one dominant satellite.

We characterise the merger orbital configuration using two angles, as shown in Fig. \ref{fig:show_nuta_prece_angle}. The satellite’s orbit relative to the main galaxy is specified by $\theta_{r,v}$, defined as the absolute value of the angle between the relative position vector $r_{\rm sat} - r_{\rm main}$ and the relative velocity vector $v_{\rm sat} - v_{\rm main}$ at $t_s$. A value of $\theta_{r,v} = 90^\circ$ corresponds to the satellite moving on a purely tangential (circular) orbit, while $\theta_{r,v} = 0^\circ$ corresponds to a head-on, purely radial orbit. 

The second angle, $\theta_{L_{\rm main}, L_{\rm orbit}}$, describes how the satellite’s orbital angular momentum, $L_{\rm orbit}$, is orientated relative to the host galaxy’s spin direction. The host’s spin, $L_{\rm main}$, is the same as defined in Section~\ref{SS:classify}. If $0<\theta_{L_{\rm main}, L_{\rm orbit}}<90$°, the satellite follows a prograde orbit aligned with the disc rotation; otherwise, its orbit is retrograde with respect to the disc spin. Both angles change over time; the evolution of $\theta_{L_{\rm main}, L_{\rm orbit}}$ will be especially substantial due to the tilt of the disc.

\subsubsection{Effects on disc tilt rate}
\label{SS:mergerorbit_effect_on_tilt_rate}

In Fig.\ref{fig:orbit_para_color_thetaw}, we show the dependence of the disc tilt rate and the disc tilt angle on the merger orbit configuration, $\theta_{\rm r,v}(t_s)$ and $\theta_{\rm L_{main},L_{orbit}}(t_s)$, for types SE and RG, respectively. 
Most type SE events that exhibit a comparatively fast tilt rate ($\omega > 10^\circ$/Gyr) are triggered by satellites on relatively tangential orbits, with $\theta_{\rm r,v} (t_s) \sim 60^\circ$, and with orbital angular momentum that is approximately vertical to the disc spin at $t_s$ ($45^\circ\lesssim  \theta_{\rm L_{main},L_{orbit}}(t_s)\lesssim 135^\circ$). These rapid tilt events eventually result in a larger overall tilt angle. 
The dependence of the tilt rate (tilt angle) on the merger angles is not obvious for type RG events. 

%------------------------------------
\subsubsection{Effects on disc tilt direction}
\label{SS:mergerorbit_effect_on_tilt_direction}
%------------------------------------
We take the spin direction of the satellite's orbit, $L_{\rm orbit}$, as our reference and use the time evolution of $\theta_{\rm L_{main},L_{orbit}}(t)$ to describe the direction of the disc tilt. We always adopt $L_{\rm orbit}$ at $t_s$ as the fixed reference; any variation in this angle is solely due to changes in the disc spin direction ($L_{\rm main}$). We quantify the tilt direction by comparing $\theta_{\rm L_{main},L_{orbit}}$ at the start of the tilt event, $t_s$, with its value at the end, $t_e$. 

In fact, the satellite’s orbital spin direction ($L_{\rm orbit}$) varies only slightly until more than $80\%$ of its mass has been disrupted, as shown in Fig.~\ref{fig:RemainMass_Lorbitchange}.  We therefore treat $L_{\rm orbit}$ as effectively unchanged between $t_s$ and $t_e$, and interpret $\theta_{\rm L_{main},L_{orbit}}(t_e)$ as representing their relative spin orientation at the end of the tilt event.

 As illustrated in Fig. \ref{fig:theta_Lmainorbit_ts_te}, the initial values of $\theta_{\rm L_{main},L_{orbit}}(t_s)$ cover nearly the full range $0$–$180^\circ$, meaning that the satellite’s orbital spin may be either prograde or retrograde with respect to the disc. In all type SE events that exhibit a large change in $\theta_{\rm L_{main},L_{orbit}}$, the tilt ends with a smaller $\theta_{\rm L_{main},L_{orbit}}(t_e)$; in other words, the disc spin consistently evolves toward the system’s stable configuration, namely alignment with the satellite’s orbital spin. 
This pattern is also seen in most type RG events, but with some notable exceptions, likely caused by a large change in $L_{\rm orbit}$ at the last stage of mergers, some of which are major mergers.

For the tilt produced by the merger of a satellite in a retrograde orbit, the disc initially has $\theta_{\rm L_{main},L_{orbit}}(t_s) > 90^\circ$ and then evolves toward zero. A wide range of angle changes is possible before the system settles into a stable configuration. Such events can typically persist for a long time and lead to large tilt angles $\Delta \theta_{\rm star}$. On the other hand, the disc will pass through an intermediate phase in which its spin is perpendicular to the satellite’s orbital angular momentum.

\begin{figure}
\centering\includegraphics[width=8cm]{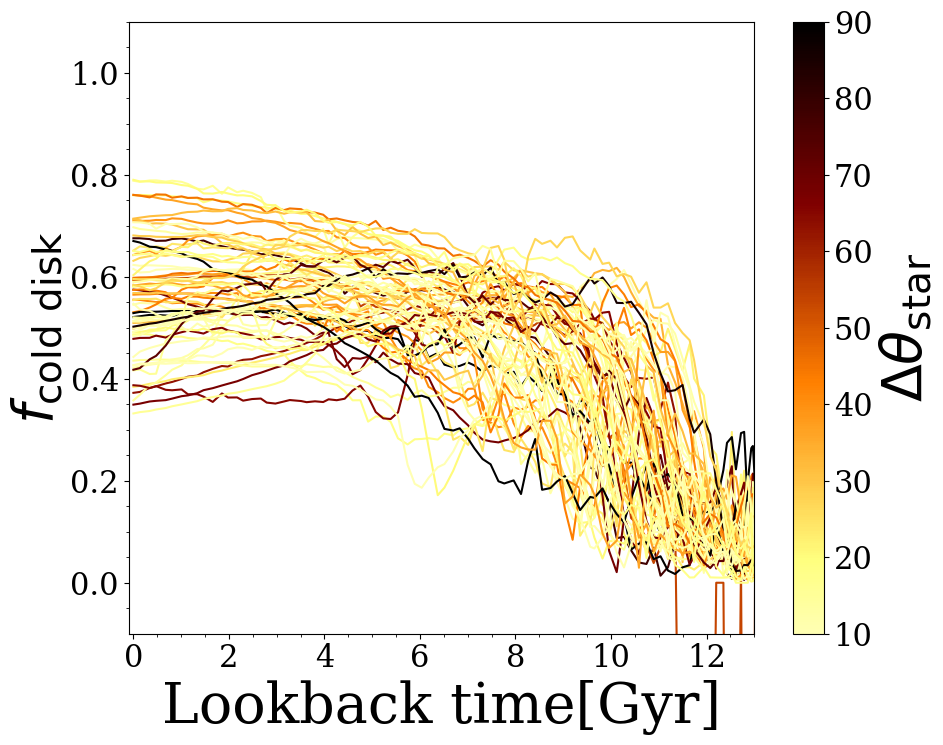}
\caption{The evolution of $f_{\rm cold\ disk}$ as a function of lookback time for the 69 galaxies with only type SE identified, each curve represents one galaxy color-coded by the overall $\Delta \theta_{\rm star}$. We cap $\Delta \theta_{\rm star}$ at $90^\circ$ to enhance color contrast for the events with high and low $\Delta \theta_{\rm star}$.}
\label{fig:fdisk_vs_time}
\end{figure}

\begin{figure*}
\centering\includegraphics[width=18cm]{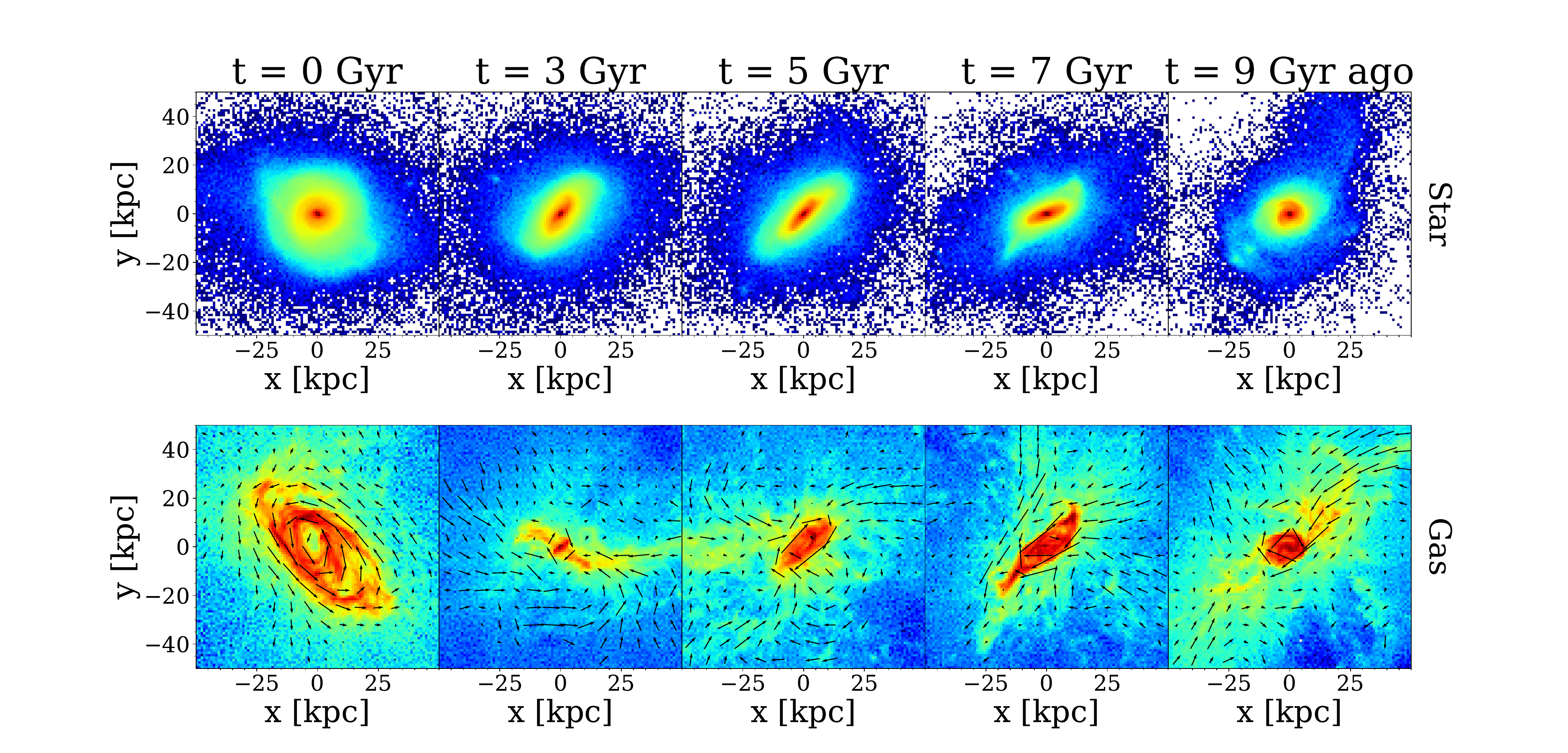}
\caption{The evolution of stellar and gaseous disks of TNG50 501208. \textbf{Top}: Stellar disk tilt as time. \textbf{Bottom}: the configuration of gas in the inner 50 kpc. The black arrows indicate the direction of motion of the gas. The gas was always set down into the stellar disk and keep aligned in the very inner regions, whereas it could also be accreted from varying directions and drop vertically into the disk.}
\label{fig:gasflow_501208}
\end{figure*}

%-------------------------------------------------------
\section{Discussion}
\label{SS:discussion}
%-------------------------------------------------------
\subsection{Effect of disk tilting process on disk evolution}
\label{SS:disk_tilt_effect_on_evolution}
%-------------------------------------------------------

We understand that the stellar disc could be significantly disrupted and regrown during a type RG event, while during the type SE event induced by small satellites, the discs are expected to evolve smoothly. To understand how Type SE events could affect disc evolution, we present the evolution of the cold disc fraction for all 69 galaxies with only type SE events identified in Fig. \ref{fig:fdisk_vs_time}. Among galaxies that undergo continuous tilting and therefore reach large tilt angles, some galaxies exhibit a decline in their cold disc fraction over the past few Gyr, likely linked to disc heating induced by minor mergers or the tilting itself. However, a substantial fraction of these systems shows a persistent increase in their cold disc fraction. Consequently, the cold disc fraction at $z=0$ for galaxies with $\Delta \theta_{\rm star} > 30^\circ$ is, from a statistical standpoint, only slightly lower than that of galaxies without a large tilt angle.

To further investigate how the disc grows when it is kept tilted, we focus on a representative galaxy, TNG50 501208, and present its evolution in Fig. \ref{fig:gasflow_501208}. Over the past few Gyr, the stellar disc has maintained a continuous tilt. We notice that it has preserved an inner thin gas disc that remains aligned with the stars. Gas accreted from large scales is initially misaligned with the inner stellar disc but could fall in along an almost vertical direction before settling into the inner disc. This process continuously supplies fuel for the ongoing growth of the stellar disc.

Disc tilting has also been proposed as an important physical mechanism for generating disc warps. However, in type SE events, we do not observe a pronounced development of a warp (see Fig. \ref{fig:warp} in the appendix). Such tilts typically evolve very slowly, with a tilt rate of only $\omega \lesssim 10^\circ$ per Gyr. The characteristic crossing time of stars in the stellar disc is about $0.1$ Gyr or less, during which the disc tilts by only about $1^\circ$. Because this crossing time is short, stars in the stellar disc can readjust rapidly and thus remain close to equilibrium. Under these conditions, the disc tilt behaves approximately as a rigid-body rotation.

%------------------------------------------------------
\subsection{Satellites association to tilt events}
\label{SS:satellites_associate_to_event}
%-------------------------------------------------------
In a cosmological setting, disc tilting can arise through several processes, such as the effects of non-spherical dark matter halo shapes, gas accretion, and merger events. Moreover, a complex merger history makes it more difficult to pinpoint the specific satellite responsible for inducing the disc tilt.

To investigate the connection between candidate satellites and tilt episodes, satellites must be identified accurately and with a reasonably high success rate; otherwise, any inferred relationship will be strongly weakened by misidentifications. We therefore first defined the onset time of each disc-tilt event ($t_s$) using the turning point in the disc’s orientation and then selected the most massive satellite associated with that onset time, which should substantially improve the likelihood of identifying the true responsible satellite. In systems where several massive satellites approach at $t_s$, focusing only on the most massive object can still lead to a substantial mis-identification risk; we therefore omit these cases from the orbital analysis.

In systems with only one massive satellite associated with the tilt event, we observe a consistent tilt direction (Fig.~\ref{fig:theta_Lmainorbit_ts_te}): nearly all discs tilt toward alignment with the spin direction of the satellite’s orbit. This coherent behaviour depends on correctly identifying the associated satellites, and we treat it as supporting evidence that the identification is indeed correct.

%---------------------------------------------------
\section{Conclusion}
\label{SS:conclusion}

We study the disc tilts of 93 Milky Way–like galaxies drawn from the cosmological simulation TNG50. We classify two categories of disc-tilt events (93 type SE and 24 type RG) with clearly defined start and end times, allowing us to quantitatively measure the duration, total tilt angle, and tilt rate of each event. Type SE events are characterised as occurring solely during phases of smooth disc evolution, without substantial disc disruption, whereas type RG events occur in conjunction with disc disruption. For each disc tilt, we also identify the most massive associated satellite galaxy and quantify its merger orbital configuration. Our main results are as follows:

\begin{itemize}

\item Over the past 9 Gyr, 64 of our 93 galaxies have experienced a net disc tilt of more than $30^\circ$. Using our definitions of type SE and type RG events, we find that 39 of the 93 type SE events and 18 of the 24 type RG events exhibit tilt angles above $30^\circ$. The type SE events can persist for extended periods, with a median duration of 6 Gyr, whereas type RG events last for shorter time periods.

\item For type SE events, both the disc tilt rate and the total tilt angle show a moderate correlation with the mass of the associated satellite. Systems with tilt angles above 30° and tilt rates greater than 10 degrees/Gyr are predominantly driven by satellites with $M_{\rm max-sat,t_s}$/$M_{\rm *, main, t_s} \sim 1$, i.e., minor mergers with total mass comparable to the galaxy's stellar mass. The type RG events are typically triggered by more massive mergers but are mostly still minor mergers.

\item The disc tilt rates are influenced by the orbital configuration of the satellite. Most type SE events that exhibit relatively rapid tilting ($\omega > 10^\circ$/Gyr or $\Delta\theta_{\rm star}>50^\circ$) are associated with satellites on tangential orbits with $\theta_{\rm r,v} \sim 60^\circ$, whose orbital angular momentum is roughly perpendicular or retrograde concerning the disc spin at the start time, i.e., with $45^\circ\lesssim\theta_{\rm L_{main},L_{orbit}}\lesssim 135^\circ$. 

\item In type SE events where a single satellite clearly dominates, almost all discs tip toward aligning with the satellite orbit’s spin direction. In particular, if the satellite arrives with a retrograde orbit, the disc’s tilt can persist for a long time and move through an intermediate phase in which it becomes nearly perpendicular to the satellite’s orbital plane.

\end{itemize}

In summary, disc tilting appears to be a frequent feature in the evolutionary histories of MW–like galaxies. Our results indicate that disc tilts induced by minor mergers are widespread; such disc tilt can persist for extended periods and permit steady growth of the dynamically cold disc. Although such a tilt will gradually evolve toward alignment with the satellite’s orbital angular momentum, the disc can remain misaligned for a long time and pass through an intermediate phase in which its spin becomes perpendicular to the satellite’s orbital angular momentum when the satellite approaches on a retrograde orbit. This mechanism has been suggested as the physical origin of the vertical alignment of the stellar disc and the plane of satellites in the MW \citep{Zhu2026A&A...706A.193Z}. Here, we show that, from a statistical standpoint, the vertical orientation is a characteristic intermediate stage in the disc’s reorientation.

%----------------------------------------------------

% ----------------------------------------------------

\begin{acknowledgement}
The research presented here is supported by the National Key R\&D Programme of China No. 2025YFF0511002, 2024YFA1611902, the NSFC key project 12633002, 12588202, 12633005, 12573024 and 12533007, CAS Project for Young Scientists in Basic Research, Grant No. YSBR-062, the Strategic Priority Research Program of Chinese Academy of Sciences grant No. XDB1160102 and the China Manned Space Programme with grant Nos. CMS-CSST-2025-A11 and CMS-CSST-2025-A20.

%The research presented here is supported by the National Key R\&D Programme of China No. 2022YFF0503403 (LZ), the NSFC key project 12633002 (LZ), and CAS Project for Young Scientists in Basic Research, Grant No. YSBR-062 (LZ). JC acknowledges the support from the NSFC projects 12573024, 12533007, and the China Manned Space Programme with grant No. CMS-CSST-2025-A20.

%X.-X.X. acknowledges the support from the National Key Research and Development Program of China No. 2024YFA1611902, National Natural Science Foundation of China (NSFC) No. 12588202 and 12633005, CAS Project for Young Scientists in Basic Research grant No. YSBR-062, the Strategic Priority Research Program of Chinese Academy of Sciences grant No. XDB1160102, and the science research grants from the China Manned Space Project with NO. CMS-CSST-2025-A11.

\end{acknowledgement}

%---------------------------------------------------------------------
\bibliographystyle{aa}  % style aa.bst
\bibliography{ms_relic} % your references Yourfile.bib
%---------------------------------------------------------------------

%---------------------------------------------------------------------
\begin{appendix}\onecolumn

\section{}

\begin{figure}
\centering\includegraphics[width=12cm]{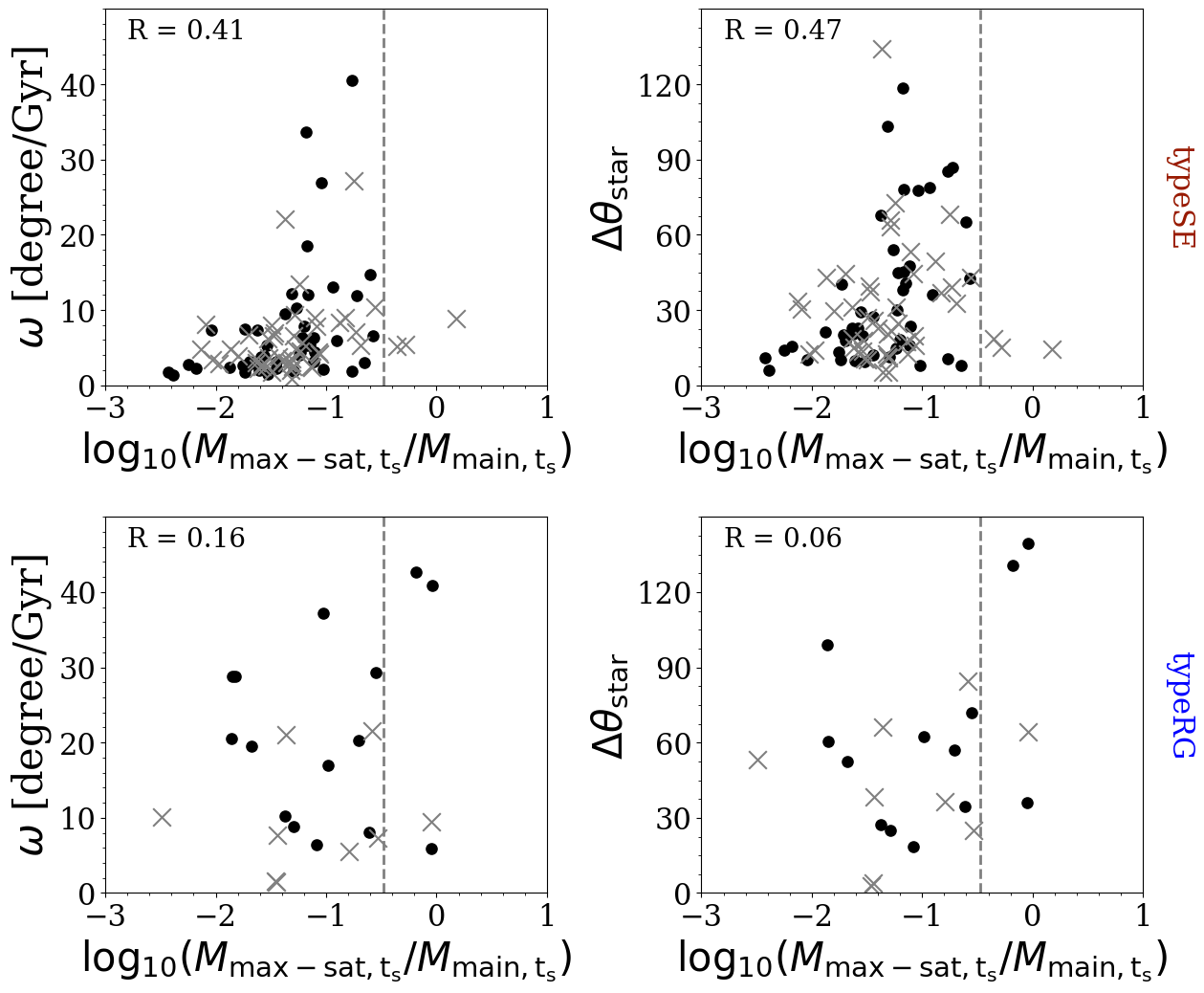}
\caption{Same as Fig.\ref{fig:mass_ratio_in_ts}, but for mass ratio between the total mass of the satellite and total mass of the host galaxy at $t_s$. The vertical lines mark the mass ratio of 1:3, which we consider as the separation of minor merger and major merger.
}
\label{fig:all_mass_ratio_in_ts}
\end{figure}

\begin{figure}
\centering\includegraphics[width=18cm]{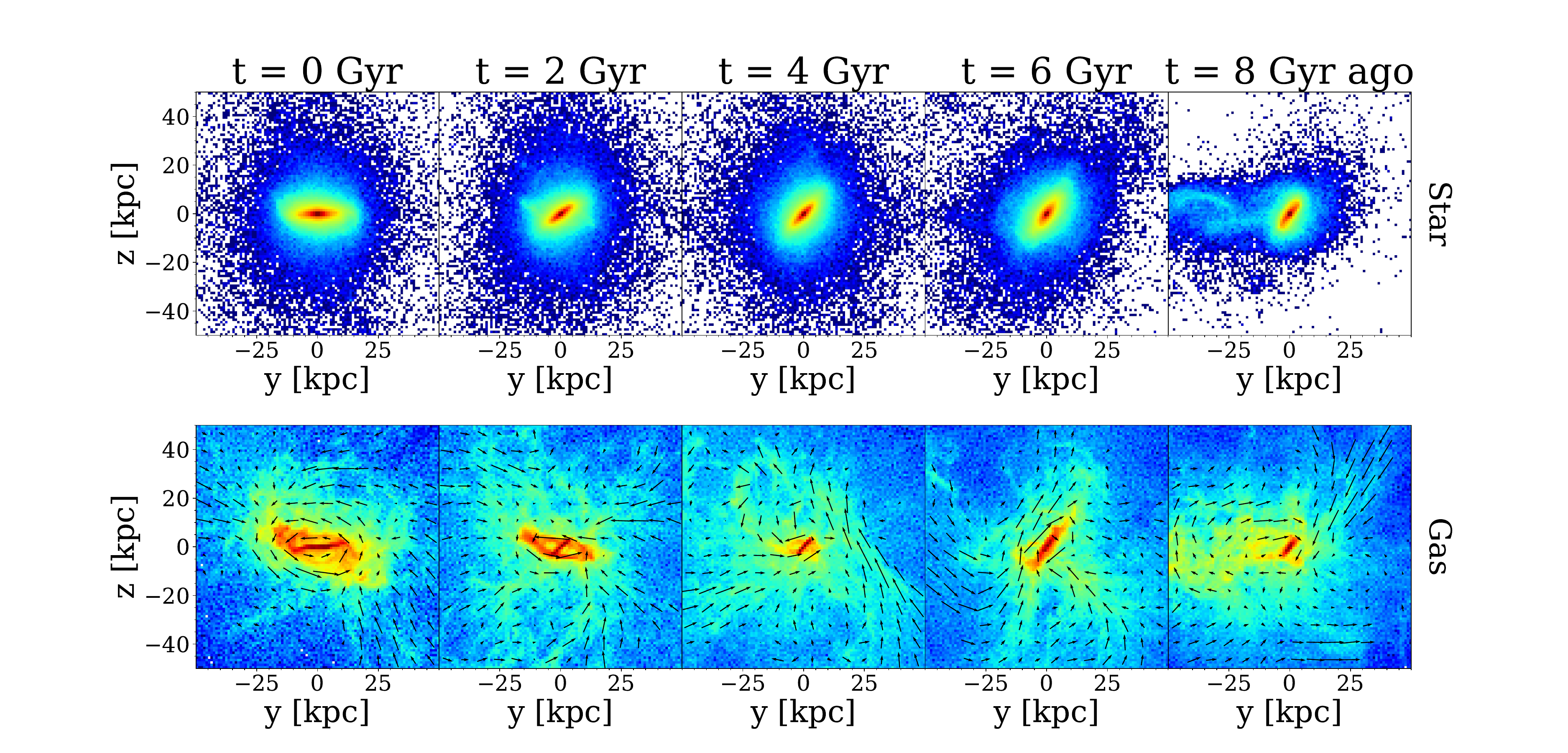}
\caption{The evolution of stellar and gaseous disks of TNG50 546474. \textbf{Top}: Stellar disk tilt as a function of time. \textbf{Bottom}: the configuration of gas in the inner 50 kpc. The black arrows indicates the direction of motion of the gas.}
\label{fig:gasflow_546474}
\end{figure}

\begin{figure}
\centering\includegraphics[width=7cm]{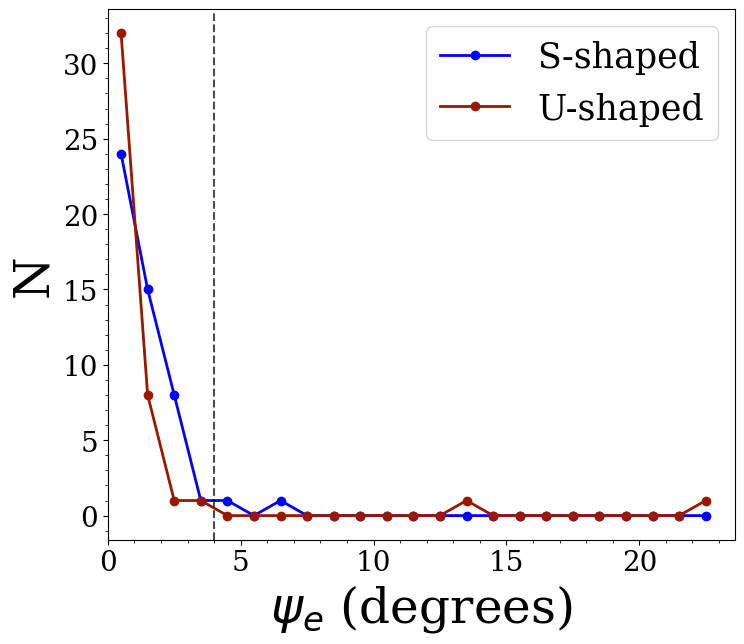}
\caption{The warp angle of all galaxies. $\Psi_{e}>4$° is defined as a strong warp. S-shaped and U-shaped are different warp shapes following the definition in \citep{Reshetnikov2025}.}
\label{fig:warp}
\end{figure}

We measured the warp angle for all 93 galaxies. Following \citet{Reshetnikov2025}, we determined the projected centreline of each disk by skeletonising its iso-mass contours: for every contour, we extracted the region it encloses and thinned it to a one-pixel-wide skeleton that follows the contour geometry. We then averaged the skeletons from the different contours to obtain the disk centreline, which would be linear if the disk were unwarped. For an S-shaped warp, the disk curves in opposite directions on the two sides of the galaxy centre, producing the familiar integral-sign morphology; for a U-shaped warp, both sides curve the same way, giving the disk a bowl-like appearance. As shown in Fig.~\ref{fig:warp}, the majority of our galaxies have $\Psi_e$ < 4°, indicating that they lack strong warps according to the criteria of \citet{Reshetnikov2025}.

\end{appendix}

\end{document}